\documentclass[a4paper]{jpconf}
\usepackage{graphicx}

\begin{document}
\title{Thermal Conductivity in the Triangular-Lattice Antiferromagnet Ba$_3$CoSb$_2$O$_9$}

\author{K Naruse$^1$, T Kawamata$^1$, M Ohno$^1$, Y Matsuoka$^1$, H Sudo$^1$, H Nagasawa$^1$, Y Hagiya$^1$, T Sasaki$^2$ and Y Koike$^1$}

\address{$^1$ Department of Applied Physics, Tohoku University, Sendai 980-8579, Japan}
\address{$^2$ Institute for Materials Reserch, Tohoku University, Sendai 980-8577, Japan}

\ead{tkawamata@teion.apph.tohoku.ac.jp}

\begin{abstract}
We have measured the thermal conductivity along the $ab$-plane and $c$-axis, $\kappa_{\rm{ab}}$ and $\kappa_{\rm{c}}$, of the $S$ = 1/2 triangular-lattice antiferromanget Ba$_3$CoSb$_2$O$_9$ single crystals in magnetic fields parallel to the $ab$-plane up to 14 T, in order to investigate the change of the magnetic state.
It has been found that $\kappa_{\rm{ab}}$ and $\kappa_{\rm{c}}$ are enhanced suddenly at low temperatures below 3.8 K, where an antiferromagnetic order of the 120$^{\circ}$ structure in the $ab$-plane occurs in zero field.
On the other hand, the magnetic-field dependence of $\kappa_{\rm{ab}}$ and $\kappa_{\rm{c}}$ at 2.8 K show increases around 8 T, where Ba$_3$CoSb$_2$O$_9$ undergoes the 1/3 magnetization plateau in magnetic fields parallel to the $ab$-plane.
Furthermore, the magnetic-field dependence of $\kappa_{\rm{ab}}$ and $\kappa_{\rm{c}}$ show dips around 13 T which is the middle of the 1/3 magnetization plateau.
These dips may be due to a possible change of the magnetic state.
\end{abstract}

\section{Introduction}
Recently, thermal conductivity in quantum spin systems has attracted considerable interst because it is closely related to the spin state.
First, large thermal conductivity due to magnetic excitations ($\kappa_{\rm spin}$) has been observed in one-dimensional quantum spin system Sr$_2$CuO$_3$ \cite{Takahashi_2006,Kawamata_2008,Sologubenko_2001} and SrCuO$_3$ \cite{Sologubenko_2001,Kawamata_2010}.
Moreover, since magnetic excitations scatter phonons, a change of magntic state affects the thermal conductivity due to phonons ($\kappa_{\rm phonon}$).
This indicates that thermal conductivity meausrements is useful probe to investigate the change of magnetic state and phase transitions.
For example, it is reported that spin gap is related to $\kappa_{\rm phonon}$ in SrCu$_2$(BO$_3$)$_2$ \cite{Kudo_2001,Hofmann_2001}.
Since the number of magnetic excitations which scatter phonons is determined by spin gap, the spin-gap formation and the reduction of spin gap lead to increase and decrease of $\kappa_{\rm phonon}$, respectively.

In Ba$_3$CoSb$_2$O$_9$, Co$^{2+}$ spins with the spin quantum number $S$ = 1/2 form an uniform triangular-lattice in the $ab$-plane.
From the analysis of ESR and magnetization processes results, intralayer antiferromagnetic interaction $J$ and interlayer antiferromagnetic interaction $J'$ are determined to be $J$ = 18.5 K and $J'$ = 0.48 K \cite{Susuki_2013}.
An antiferromagnetic order of the 120$^{\circ}$ structure in the $ab$-plane occurs at $T_{\rm N}$ $\approx$ 3.8 K due to weak interlayer interaction $J'$ in zero field \cite{Shirata_2012}.
When the magentic field is applied along the $ab$-plane, the up-up-down (UUD) structure appears in the magnetic fields above 8 T along the $ab$-plane.
This state causes a 1/3 magnetization plateau.
On the other hand, no magnetization plateau is observed and a cusp appears for the magnetic field along $c$-axis.
These results indicate the easy-plane anisotropy, which is consistent with the collective ESR modes \cite{Susuki_2013}.
Magnetic structure of Ba$_3$CoSb$_2$O$_9$ in the magnetic fields has been also investigated \cite{Susuki_2013,Koutroulakis_2013}.

It is predicted theoretically that quantum fluctuation stabilize the UUD structure in triagular-lattice antiferromagnets \cite{Chubukov_1991}.
Experimentally, a 1/3 magnetization plateau in the $S$ = 1/2 triangular-lattice antiferromagnet (TLAF) Cs$_2$CuBr$_4$ has been investigated \cite{Ono_2003,Ono_2004,Fortune_2009}.
Since an magnetization plateau is caused by an enegy gap which separates low-energy excitaions from the ground state, the UUD state in Cs$_2$CuBr$_4$ is gapped state \cite{Tsujii_2007,Fujii_2007}.
Thermal conductivity is a useful probe to investigate the change of the magnetic state and the spin gap which stabilizes a 1/3 magnetization plateau.
Moreover, $\kappa_{\rm spin}$ in triganular-lattice antiferromagnets is already investigated in $\kappa$-(BEDT-TTF)$_2$Cu$_2$(CN)$_3$ \cite{Yamashita_2009}.
In order to investigate the existance of $\kappa_{\rm spin}$ and the change of the magnetic state in the UUD state with the 1/3 magnetization plateau we have measured the thermal conductivity of Ba$_3$CoSb$_2$O$_9$ in magentic fields.

\section{Experimental}
Single crystals of Ba$_3$CoSb$_2$O$_9$ were grown by floating-zone method.
The quality of the single crystals was checked by the x-ray back-Laue photography to be good.
Thermal conductivity measurements were carried out by the conventional steady-state method.
Magnetic fields up to 14 T were applied parallel to the $ab$-plane.

\section{Results and discussion} 
Figure 1 shows the tempetature dependence of the thermal conductivity along the $ab$-plane and $c$-axis, $\kappa_{\rm ab}$ and $\kappa_{\rm c}$, for single crystals of Ba$_3$CoSb$_2$O$_9$.
$\kappa_{\rm ab}$ and $\kappa_{\rm c}$ show a broad peak around 40 K.
While $\kappa_{\rm ab}$ and $\kappa_{\rm c}$ decrease with decreasing temperature below 40 K, they increase below $T_{\rm N}$ as shown in the inset of Fig. 1.
Although temperature dependence of $\kappa_{\rm phonon}$ usually shows a peak around 10 K, it was not observed not around 10 K but around 40 K in Ba$_3$CoSb$_2$O$_9$.
Magnetic fluctuation arising from frustration may suppress the mean free path of phonons ($l_{\rm phonon}$). 
Since the absolute values of the peaks between $\kappa_{\rm ab}$ and $\kappa_{\rm c}$ is much same and the behaviour of the temperature dependences of $\kappa_{\rm ab}$ and $\kappa_{\rm c}$ is isotropic, phonon contribution is dominant in the thermal conductivity.
Enhancemenet of $\kappa_{\rm ab}$ and $\kappa_{\rm c}$ at the temperature below $T_{\rm N}$ indicates that $l_{\rm phonon}$ is extended due to the suppression of phonon scattering in the ordered state.

\begin{figure}[t]
\centering
\includegraphics[width=0.45\linewidth]{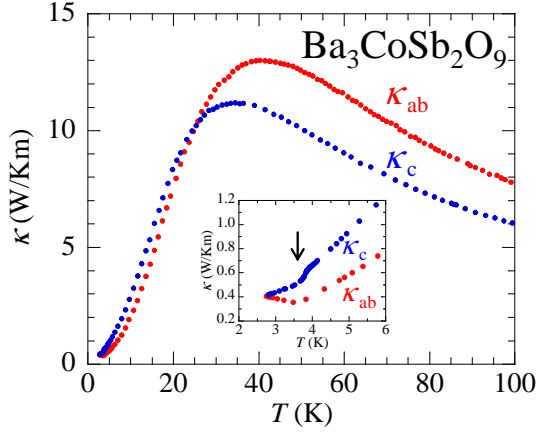}
\hspace{2pc}
\begin{minipage}[b]{.4\textwidth}
\centering
\caption{Tempetature depencence of the thermal conductivity along the $ab$-plane and $c$-axis, $\kappa_{\rm ab}$ and $\kappa_{\rm c}$, for single crystals of Ba$_3$CoSb$_2$O$_9$.
The inset shows an expansion of the graph around $T_{\rm N}$ = 3.8K.}
\label{fig01}
\end{minipage}
\end{figure}

\begin{figure}[t]
\centering
\includegraphics[width=0.8\linewidth]{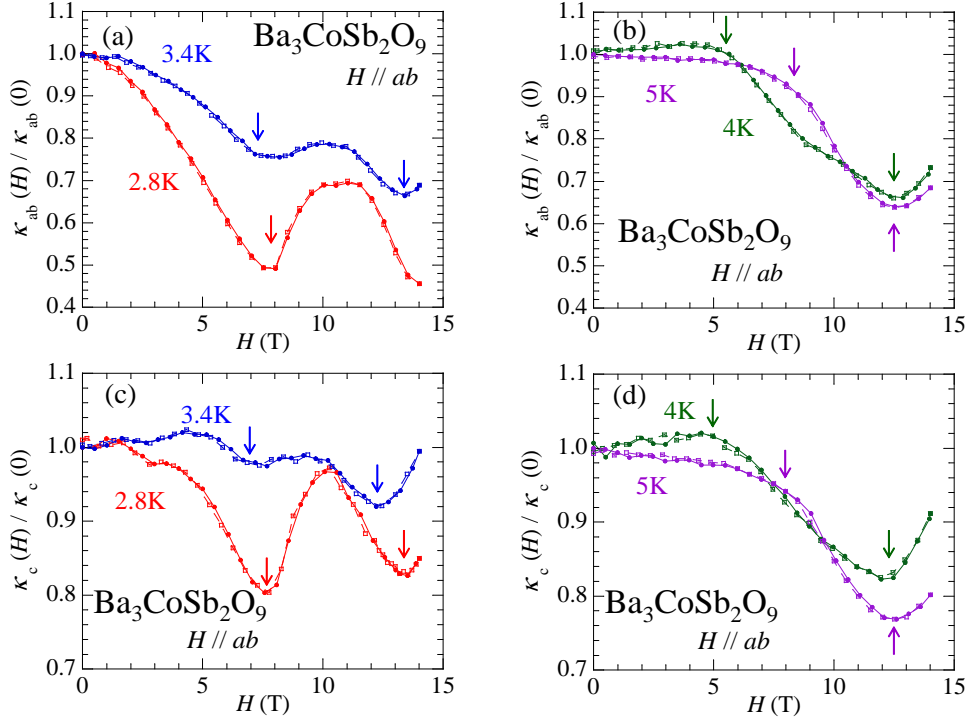}
\caption{Field depencence of $\kappa_{\rm ab}$ and $\kappa_{\rm c}$, $\kappa_{\rm ab}(H)$ and $\kappa_{\rm c}(H)$, (normalized to the 0 T value) for single crystals of Ba$_3$CoSb$_2$O$_9$ at 2.8 K, 3.4 K, 4 K and 5 K.}
\label{fig02}
\end{figure}

Figure \ref{fig02} shows the field dependences of $\kappa_{\rm ab}$ and $\kappa_{\rm c}$, $\kappa_{\rm ab}(H)$ and $\kappa_{\rm c}(H)$, at 2.8 K, 3.4 K, 4 K and 5 K.
There is no difference between $\kappa_{\rm ab}(H)$ and $\kappa_{\rm c}(H)$, which is consistent with the dominant phonon contribution to the thermal conductivity in Ba$_3$CoSb$_2$O$_9$.
Since phonons can be scattered by magnetic excitations, the behaviour of $\kappa_{\rm ab}(H)$ and $\kappa_{\rm c}(H)$ attributes the change of the magnetic state.
$\kappa_{\rm ab}$ and $\kappa_{\rm c}$ decreased with increasing magnetic field in the 120$^{\circ}$ structure phase at 2.8 K and 3.4 K, which are below $T_{\rm N}$.
This decrease is due to the suppression of $l_{\rm phonon}$, because the number of magnetic excitations increases with increasing magnetic field.
$\kappa_{\rm ab}$ and $\kappa_{\rm c}$ increased suddenly around 8 T where Ba$_3$CoSb$_2$O$_9$ undergoes the transition from the 120$^{\circ}$ structure phase to the UUD phase.
This increase may indicate that the number of magnetic excitations decreased due to the appearance of the spin gap in the UUD phase and that suppressed $l_{\rm phonon}$ increased.
$\kappa_{\rm ab}(H)$ and $\kappa_{\rm c}(H)$ at 4 K and 5 K show no change in low magnetic field as shown in Fig. \ref{fig02}(b,d), because these temperatures are in the paramagnetic (PM) phase.
Then $\kappa_{\rm ab}$ and $\kappa_{\rm c}$ decreased due to the transition from the PM phase to the UUD phase at 5 T and 7 T for 4 K and 5 K, respectively.

\begin{figure}[t]
\centering
\includegraphics[width=0.45\linewidth]{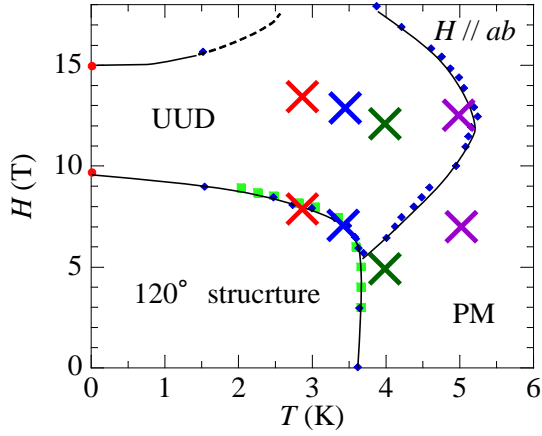}
\hspace{2pc}%
\begin{minipage}[b]{.4\textwidth}
\centering
\caption{$H$--$T$ magnetic phase diagram of Ba$_3$CoSb$_2$O$_9$ in magnetic fields along the $ab$-plane. Data were taken from ref. \cite{Koutroulakis_2013,Zhou_2012}. $\times$ marks are placed at the change in the field dependence of the thermal conductivity shown as the arrows in Fig. \ref{fig02}.}
\label{fig03}
\end{minipage}
\end{figure}

While $\kappa_{\rm ab}$ and $\kappa_{\rm c}$ increased at the transition from the 120$^{\circ}$ structure phase to the UUD phase (Fig. \ref{fig02}(a,c)), $\kappa_{\rm ab}$ and $\kappa_{\rm c}$ decreased at the transition from the PM phase to the UUD phase (Fig. \ref{fig02}(b,d)).
This difference may be explaned by the number of magentic excitations.
In the 120$^{\circ}$ structure phase, the number of magentic excitations is increased due to the magnetic field.
At the transition from the 120$^{\circ}$ structure phase to the UUD phase, the number of magnetic excitations decreases due to the appearance of the spin gap, which leads to the extension of $l_{\rm phonon}$.
On the other hand, in the PM phase, the number of magnetic excitations do not increase with increasing magnetic field.
At the transition from the PM structure phase to the UUD phase, the number of magnetic excitations may increase and $l_{\rm phonon}$ may be suppressed.
Since the thermal fluctuation is large near the phase boundary in the UUD phase, the extension of $l_{\rm phonon}$ is not observed even in the existance of the spin gap.
Arrows in Fig. \ref{fig02} indicate transition to the UUD phase and the transition fields are shown as $\times$ marks in Fig. \ref{fig03}.
Transition fields from the 120$^{\circ}$ structure phase to the UUD phase are much same with the magnetic phase diagram of ref. \cite{Koutroulakis_2013,Zhou_2012} shown in Fig. \ref{fig03}.
On the other hand, the transition fields from the PM phase to the UUD phase are different.
This difference suggests that the magnetic fluctuation appears at higher temperature than the phase boundary by $\sim$1 K in the PM phase.
Existance of the magnetic fluctuation coincides with the broad peak in the temperature dependence of the specific heat at the transition from the PM phase to the UUD phase \cite{Zhou_2012}.

In high magnetic field above 10 T, $\kappa_{\rm ab}(H)$ and $\kappa_{\rm c}(H)$ show peaks around 11 T at 2.8 K and 3.4 K (Fig. \ref{fig02}(a,c)).
Although these peaks may indicate that the magnitude of the spin gap starts decreasing with increasing magnetic field around around 11 T, the origin of these peaks is unclear.
Futhermore, $\kappa_{\rm ab}(H)$ and $\kappa_{\rm c}(H)$ show dips around 13 T at all temperatures 2.8 K, 3.4 K, 4 K and 5 K shown as arrows in Fig. \ref{fig02}.
Arrows in Fig. \ref{fig02} indicate the dips around 13 T and the fields where the dips appear are shown as $\times$ marks in Fig. \ref{fig03}.
Figure \ref{fig03} shows that these dips are placed in the center of the UUD phase.
However, no transition is observed in other measurements.
Although the field dependences of the specific heat do not show any anomalies which indicate phase transition around 13 T \cite{Zhou_2012}, some change may occur.
It is necessary to investigate this magentic state using NMR and neutron scattering measurements.

\section{Summary}
In order to investigate the existance of $\kappa_{\rm spin}$ and the change of the magnetic state, we have measured the thermal conductivity along the $ab$-plane and $c$-axis, $\kappa_{\rm{ab}}$ and $\kappa_{\rm{c}}$, the $S$ = 1/2 triangular-lattice antiferromanget Ba$_3$CoSb$_2$O$_9$ single crystals in magnetic fields.
It has been found that the phonon contribution to the thermal conductivity is dominant due to the isotropic behaviour of the temperature dependence between $\kappa_{\rm{ab}}$ and $\kappa_{\rm{c}}$.
Moreover, $\kappa_{\rm{ab}}$ and $\kappa_{\rm{c}}$ increased at the temperature below $T_{\rm N}$, because the decrease of the number of magnetic excitations leads to the extension of $l_{\rm phonon}$.
In magnetic field along $ab$-plane at 2.8 K, $\kappa_{\rm{ab}}$ and $\kappa_{\rm{c}}$ increased around 8 T, where Ba$_3$CoSb$_2$O$_9$ undergoes the UUD phase.
This enhancement may be due to the appearance of the spin gap.
Furthermore, the magnetic-field dependence of $\kappa_{\rm{ab}}$ and $\kappa_{\rm{c}}$ show dips around 13 T which is roughly the midpoint of the 1/3 magnetization plateau.
These dips may be due to a possible change of the magnetic state.

\section*{Acknowledgments}
The thermal conductivity measurements were performed at the High Field Laboratory for Superconducting Materials, Institute for Materials Research, Tohoku University.
This work was supported by a Grant-in-Aid for Scientific Reserach from the Ministry of Educcation, Culture, Sports, Science and Technology, Japan.

\end{document}